\documentclass[journal]{vgtc}                     

\onlineid{0}

\vgtccategory{Research}

\vgtcpapertype{application/design study}

\title{DeLVE into Earth’s Past: A Visualization-Based Exhibit Deployed Across Multiple Museum Contexts}

\author{
  \authororcid{Mara Solen}{0000-0002-5191-5193},
  \authororcid{Nigar Sultana}{0009-0001-6639-7034}
  \authororcid{Laura Lukes}{0000-0003-3648-0942},
  \authororcid{Tamara Munzner}{0000-0002-3294-3869}
}

\authorfooter{
    \item Mara Solen and Tamara Munzner are with the University of British Columbia, Department of Computer Science. E-mail: \{solen,tmm\}@cs.ubc.ca.
    \item Nigar Sultana and Laura Lukes are with the University of British Columbia, Department of Earth, Ocean, and Atmospheric Sciences. E-mail: \{nsultana,llukes\}@eoas.ubc.ca.
}

\abstract{While previous work has found success in deploying visualizations as museum exhibits, \change{it has not investigated whether museum context impacts visitor behaviour with these exhibits}. We present an interactive Deep-time Literacy Visualization Exhibit (DeLVE) to help museum visitors understand deep time (lengths of extremely long geological processes) by improving proportional reasoning skills through comparison of different time periods. DeLVE uses a new visualization idiom, Connected Multi-Tier Ranges, to visualize curated datasets of past events across multiple scales of time, relating extreme scales with concrete scales that have more familiar magnitudes and units. Museum staff at three separate museums approved the deployment of DeLVE as a digital kiosk, and devoted time to curating a unique dataset in each of them. We collect data from two sources, an observational study and system trace logs\delete{, yielding evidence of successfully meeting our requirements}. We discuss the importance of context: similar museum exhibits in different contexts were received very differently by visitors. We additionally discuss differences in our process from \change{Sedlmair et al.'s} design study methodology which is focused on design studies \change{triggered by connection with collaborators rather than the discovery of a concept to communicate}. Supplemental materials are available at:~\url{https://osf.io/z53dq/}}


\teaser{
  \centering
  \includegraphics[width=0.9\textwidth]{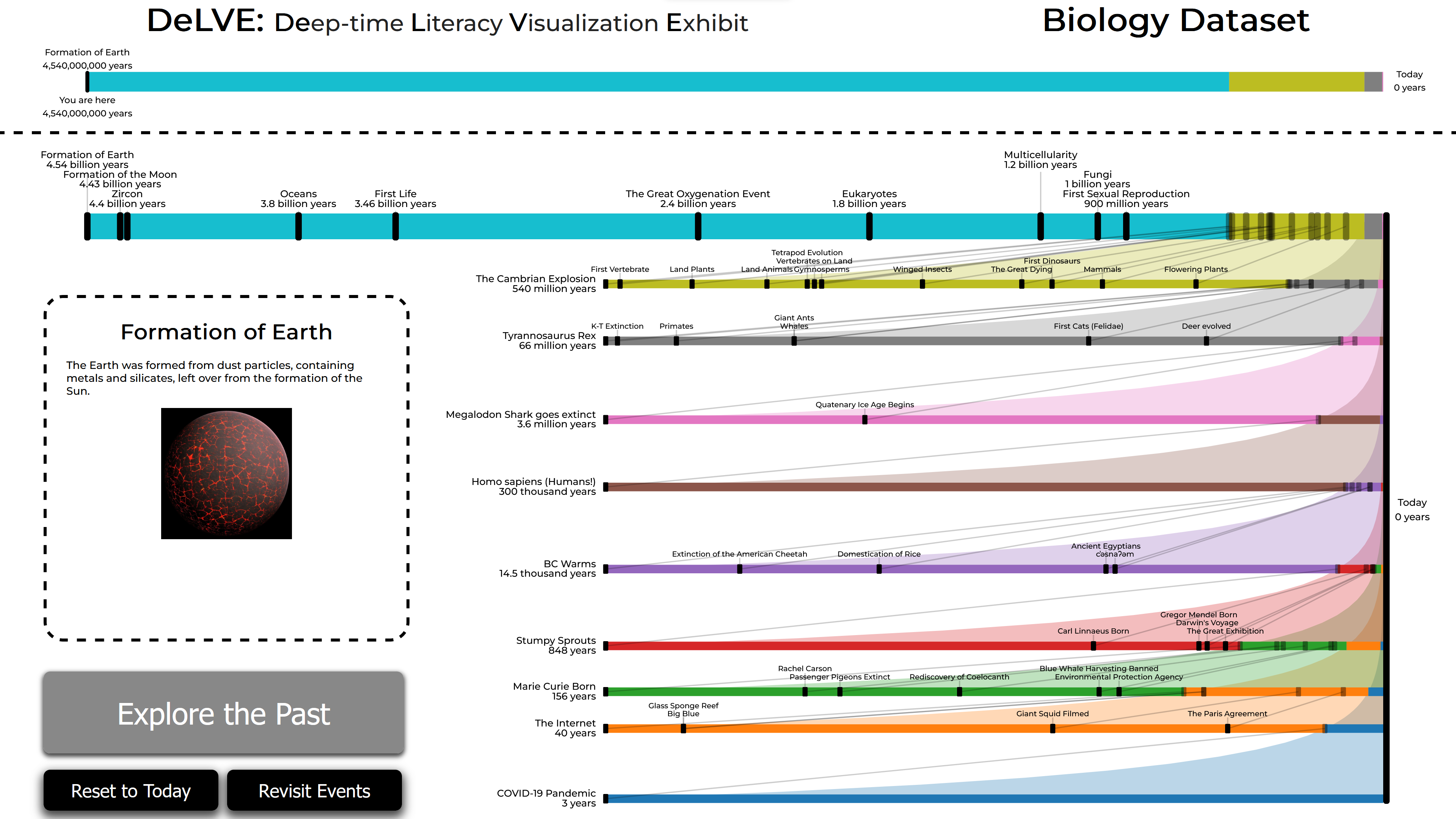}
  \caption{The interface of DeLVE, after progressing through a full dataset on biology.}
  \label{fig:delve-overview-teaser}
}

\graphicspath{{figs/}{figures/}{pictures/}{images/}{./}} 

\usepackage{tabu}                      
\usepackage{booktabs}                  
\usepackage{lipsum}                    
\usepackage{mwe}                       

\usepackage{mathptmx}                  

\usepackage[normalem]{ulem}
\usepackage{subcaption}

\begin{document}

\newcommand{\tmC}[1]{\textcolor{purple}{TM: #1}}
\newcommand{\msC}[1]{\textcolor{blue}{MS: #1}}

\definecolor{ChangedContentColor}{HTML}{C7372F}  
\newenvironment{NEW_ENV}{\color{ChangedContentColor}}{}


\newcommand{\change}[1]{{#1}}
\newcommand{\delete}[1]{}


\newcommand{\suppRef}[1]{Supp.~{\S}{#1}}

\newcommand{\SuppSystem}[0]{\suppRef{2}}
\newcommand{\SuppSystemEvol}[0]{\suppRef{2.1}}
\newcommand{\SuppDeployments}[0]{\suppRef{3}}
\newcommand{\SuppResultFigs}[0]{\suppRef{4}}
\newcommand{\SuppMethods}[0]{\suppRef{5}}


\firstsection{Introduction}

\maketitle

Digital visualizations can make for effective and engaging museum exhibits for many reasons including the potential for interactivity and their ability to expose large audiences to scientific datasets \cite{dasuMMF2020, maLMF2012}. However, designing such interactions is still a challenge, particularly when considering the diversity of museum contexts. We define a \textbf{museum context} as a specific physical space within a specific museum, with all concomitant attributes of how it is experienced by visitors including its theme, style, size, intended use, exhibits, and the audience who visits that space. Different rooms within a single museum may entail different contexts. Given the differences in user behaviour documented by O'Reilly and Inkpen in varying visualization study environments \cite{reillyI2007}, we hypothesize that museum context will also play an important role in visitor behaviour. 

In this design study, we present a Deep-time Literacy Visualization Exhibit (\textbf{DeLVE}). DeLVE is an interactive tool that museum visitors can use to explore past events across different scales of time, designed with the intention of improving visitors' sense of \textbf{deep time}: a geoscientist term for the very long periods of time of geological processes \cite{zen2001}. In particular, we designed DeLVE to promote proportional reasoning, one important skill associated with deep time, through comparisons of the different scales. DeLVE addresses five requirements we identified through consultation with museum staff. Two are \change{high-level and} museum-internal: Deploy and Curate. \change{Two are high-level and visitor-facing: Engage and Inspire. One is specific to our educational focus of proportional reasoning and visitor-facing: Compare.}

While we did not begin the project with deployment agreements, we developed collaborative relationships early on with three local museums and successfully deployed an instance of DeLVE in each of them. At the time of submission, DeLVE has been deployed for over one year, and two of our collaborating museums have committed to long-term deployment and are working to find final locations, datasets, and design signage to support the exhibit. \change{We conducted an observational study in these three institutions and collected system trace logs. Our preliminary analysis of the data from these two sources suggests that museum context impacts visitor behaviour, and could inform future work which studies the impact of differences in environment or audience in more depth.} Finally, we reflect on the project process, and discuss differences between standard visualization design study methodology and that for presentation-focused design studies.

We present three primary contributions. First, the task abstraction, design requirements, and data abstraction for learning about deep time in museums. Second, the design and development of the DeLVE museum exhibit. Third, the results of and \delete{generalizeable} reflections from the deployment of DeLVE in four museum contexts across three institutions, based on the analysis of observations and system logs from visitor usage.

We also provide two secondary contributions: a visualization technique for visualizing data with quantitative measures, such as past events, on multiple scales, and an extension of \change{Sedlmair et al.'s} design study methodology \cite{sedlmairMM2012} to \change{concept-first design studies}.

\change{We do not validate DeLVE or our proposed visualization technique's task, design, or long-term educational impact. Given our opportunity to deploy in multiple institutions, we instead choose to investigate whether museum context impacts visitor behaviour. Other assessments are beyong the scope of this paper.}

\section{Related Work}
\label{sec:related-work}

We now discuss related work, divided into
visualizations as museum exhibits, deep time education in formal settings, visualizations of multiscale data for presentation, and design study methodologies.

\subsection{Visualizations as Museum Exhibits}
\label{sec:related-work:visualizations-as-museum-exhibits}

We discuss related work from other visualization research conducted in museums. Designing museum exhibit visualizations brings many challenges that are different from designing for data analysts or other professionals. Existing literature in this domain include design studies \cite{dasuMMF2020, hinrichsSC2008, maLMF2012, blockHPDES2012, viegasPHD2004} and empirical studies \cite{maMF2019, hornPEBDS2016, bornerMBH2016, peppler2021cultivating}, many of which contribute important design considerations, challenges, and principles.

Many museum visitors are personally motivated to engage with exhibits. In order to promote meaningful engagement, designs must both attract and sustain visitors \cite{dasuMMF2020, hinrichsSC2008} within ten seconds \cite{horneckerS2006}. Visitors are attracted to exhibits through entry points, after which they may choose to engage deeply. For digital exhibits, the use of special entry point screens that are disconnected from the educational content is discouraged, and the merging of visually-interesting and engaging components with educational content is beneficial \cite{blockHPDES2012}. \change{Additionally, making the entry points personally meaningful further supports engagement \cite{peppler2021cultivating}.}

Length and type of engagement with museum exhibits is diverse. In some cases, visitors may engage for a very short time, so designs must be fast to decode, interpret, and gain value from \cite{dasuMMF2020, hinrichsSC2008, bornerMBH2016}. Designers also need to prioritize which data and functionality is given priority so that short interactions can still have potential for teaching key concepts \cite{maLMF2012, blockHPDES2012}. Length of engagement can also be long, indicating deeper exploration, in which case visitors should be rewarded with additional insight \cite{hinrichsSC2008}. While some visitors may be open to exploring large datasets, others may become overwhelmed and discouraged, so it is important to support both exploratory and guided engagement styles \cite{hinrichsSC2008, maLMF2012}.

Museum visitors are a diverse group of people. They could be anywhere from novices to experts, so designs should not assume prior knowledge but instead provide information when necessary \cite{dasuMMF2020, maLMF2012}. Users can also explore museums in groups and designs should support multiple users in viewing and interacting with the exhibit \cite{dasuMMF2020, hinrichsSC2008, blockHPDES2012}. Finally, over-use of complex scientific data and over-emphasis on accuracy can lead to misconceptions; careful selection of key concepts and visual simplicity can be beneficial for visitor learning \cite{maLMF2012, blockHPDES2012}.

The learning goals of museum visualizations in existing literature primarily focus on the specific datasets they visualize \cite{dasuMMF2020, maLMF2012, viegasPHD2004, hinrichsSC2008, blockHPDES2012}. 
However, our higher-level learning goal of promoting proportional reasoning does not rely on any specific dataset; it can be enabled by any dataset which follows the data abstraction presented in Section~\ref{sec:data-abstraction}. This objective differs from design studies in the existing literature, which do not describe specific high-level learning goals for museum visitors who interact with their exhibits.

In addition, previous museum visualization studies deploy their exhibits in single institutions. Through multiple deployments, we hope to better understand how varying museum contexts affect visitor behaviour with a common exhibit.

\begin{figure*}[t]
  \centerline{\includegraphics[width=.8\textwidth]{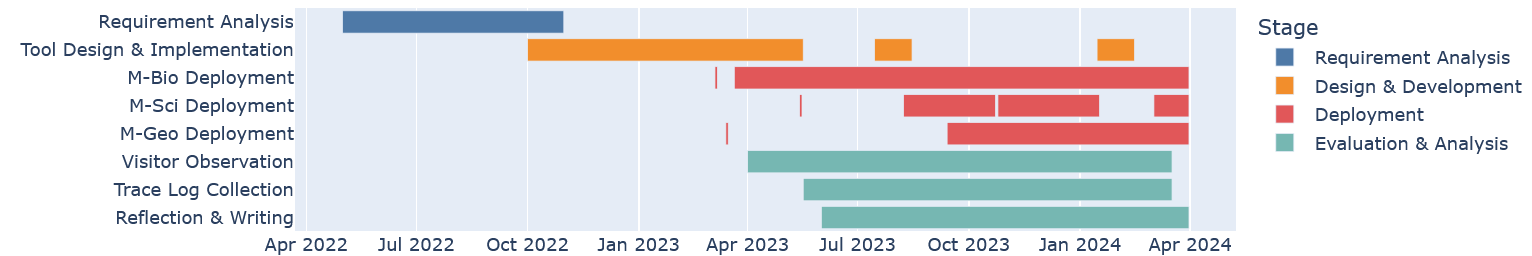}}
  \caption{DeLVE project timeline, broken down according to its four stages.}
  \label{fig:overall-gantt}
\end{figure*}

\subsection{Deep Time Formal Education}
\label{sec:related-work:deep-time-formal-education}

Many geology courses teach concepts that require an understanding of deep time. Because of this requirement, it can be included, sometimes even implicitly, in introductory geology, historical geology, structural geology, geomorphology, and geology field work classes \cite{czajkaM2018}. 
The geoscience education research literature also includes specific learning exercises for understanding deep time through proportional reasoning, such as one activity where students interact with an increasingly complex visual representation of time-based data \cite{dodickO2006} and 
one from Resnick et al.~where students continuously map larger and larger time periods to the same physical space while indicating where previous time periods appear \cite{resnickDNS2017}. While these deep time learning techniques show evidence of success in teaching students about deep time, they take too long to incorporate into a museum exhibit: the time required ranges from many minutes to many hours, while in contrast we aim for visitors to interact with the exhibit for between one and five minutes. Despite Resnick et al.~being infeasible as a museum exhibit, we are inspired by their approach and design DeLVE to conduct a similar, but faster, exercise in a digital exhibit.

\subsection{Visualizations of Multiscale Data for Presentation}

Our work is informed by the design space and high-level strategies for visualizations with large scale-item ratio proposed by Anonymized et al.~\cite{solen2024designspace}, which analyzes a collection of 54 such examples 
covering both analysis and presentation use cases, drawn from both academic literature and real-world use. In DeLVE, we use the strategy they call \textit{Familiar Zoom}, which entails zooming through a series of scales that include at least one familiar, concrete scale.

DeLVE's design uses concrete scales, 
a technique for helping people understand unfamiliar measures by relating them to familiar measures \cite{nieman2011concrete}, to support museum visitors in relating to and understanding deep time.   We were inspired by Chevalier et al.'s taxonomy of the object types and measure relations involved with concrete scales, and their set of strategies for using concrete scales \cite{chevalierVG2013}.

\subsection{Design Study Methodologies}

Sedlmair et al.'s design study methodology is a nine-stage process model that identifies specific stakeholder roles \cite{sedlmairMM2012}, but their
general-case methodology does not adequately suit all design studies, leading researchers to develop adaptations of it for specific scenarios. Syeda et al.~constructed a design study methodology for expedited design studies, specifically to support the teaching of visualization \cite{syeda2020design}. Oppermann et al.~constructed a version for data-first design studies, where the study is prompted by the acquisition of data rather than stakeholder analysis questions \cite{oppermann2020data}. 
However, previous work does not address how to conduct design studies for presentation-focused scenarios; we address that gap in this paper. 

\section{Process}
\label{sec:process}

This project took place from May 2022 to March 2024. The four authors of this paper make up the \textbf{design team}, which is an internal collaboration between a two-person visualization \textbf{(vis) team} and a geoscience education researcher \textbf{(GER) team} working at the same institution, who initiated this project on educating museum visitors about deep time. We also work with the external collaborators of museum staff at three museums in a North American city: two on-campus museums focused on geoscience and biology respectively, and one general science centre in the city core. We worked with multiple staff from each museum, primarily for understanding their goals, gaining their feedback on designs, and deploying our exhibit.

We divide the project into four stages, as shown in Figure~\ref{fig:overall-gantt}: requirement analysis and abstraction, design and development, deployment, and evaluation and analysis. The requirement analysis and abstraction stage involved literature review and interviews with museum staff to collect requirements, and multiple rounds of reflective synthesis to refine them. The design and development stage involved weekly design and prototype iteration. We succeeded in deploying DeLVE in three locations. Deployment happened differently at each of the three museums, but always started with a deployment approval meeting and involved continuous communication between the design team and the museum staff. At each of the venues, museum staff curated content to display that connected to their collections. During the evaluation and analysis stage, we conducted a study of museum visitors as part of our evaluation, drawing from two sources: direct observation of visitors, and analysis of  system trace logs. We then reflected on the entire project. 
We discuss the process details of each of these four stages in Sections~\ref{sec:requirements-and-abstractions},~\ref{sec:design},~\ref{sec:deployment}, and~\ref{sec:evaluation} respectively. 

\section{Requirements and Abstractions}
\label{sec:requirements-and-abstractions}

We present our requirement collection and analysis methods,  task abstraction,  requirements, and our data abstraction.

\subsection{Methods}

We began with ideation, as we had begun the project with the broad goal of making a visualization-based museum exhibit related to the concept of deep time. We eventually narrowed our scope to a foundational aspect of deep time: reasoning about numbers at varying and often extremely large magnitudes, a skill called \textbf{proportional reasoning}.

We then conducted remote semi-structured interviews with museum staff at local museums to better understand the internal processes and goals of their institutions, get feedback on project ideas and assess the level of buy-in for collaboration, and narrow down our project scope. In contrast to previous visualization design studies focused on museums, we did not start the project with an agreement to work with a partner organization, requiring assessment of whether institutions were interested in collaboration at an intermediate stage. We conducted four expert interviews in total, each of them around one to one and a half hours, each with \change{one or two museum staff}. There were six total participants, two from each \change{of our three collaborating museums}, all recruited through existing connections from the GER team.
We provide the interview script and transcripts in supplemental materials. 
All three museums indicated strong interest in collaboration, leading us to continue with all three and study differences in visitor behaviour across multiple deployments in different museum contexts.

From the interviews, we collected a list of museum staff goals. Some of them, such as providing visitor access to specimens, were infeasible for us based on our resources and expertise, and we deemed them to be out of scope. After determining which goals were in scope, we framed them as requirements for our design.

\subsection{Task Abstraction}

While ideating, we identified many potential tasks involving deep time, including understanding big numbers \cite{cheek2012, resnickDNS2017}; orderings, timings, and causalities of events from Earth's past \cite{czajkaM2018, libarkinKA2007, zen2001, karlstromSCPGDWHCW2008, dodickO2006}; and how rates of change affect geological processes \cite{czajkaM2018, jolleyJH2013}. 

As we narrowed our focus to supporting proportional reasoning, we investigated existing methods for teaching the concept. We identified Resnick et al.'s classroom exercise, described in Section~\ref{sec:related-work}, as foundational inspiration \cite{resnickDNS2017}.
While the original activity is done using physical materials, we aim to facilitate a faster version of the experience through digital interaction. This existing literature indicates that having learners make comparisons between time periods of different magnitudes can help with learning proportional reasoning. We thus focus on anchoring and situating relatable time scales to more extreme ones as an appropriate focus for in this exhibit. Accordingly, we selected \textbf{compare varied-magnitude time periods} as the primary task.

\subsection{Design Requirements}

We identified five requirements to address through our design: Deploy, Curate, Engage, Inspire, and Compare. The first two requirements are museum-internal requirements and the final three are visitor-facing requirements. The first four are \change{high-level} requirements that pertain to most museum contexts; the last is more specific to our situation. \change{We omit lower-level details from our high-level requirements to retain generalizability.}

\subsubsection{R-Deploy}

Deployment of a research project within a museum requires buy-in from the staff at that institution. Our first requirement is to find and build collaborations with staff at local museums both to understand their needs and to create an exhibit they feel comfortable deploying for their visitors (R-Deploy). \delete{Through the expert interviews, we found that museum staff at all three museums were very enthusiastic about collaboration. These domain experts immediately offered to devote further time to talking with us in support of the project, and advised us that there were unlikely to be barriers to eventual deployment so long as we communicated sufficiently.
This outcome stood in contrast to the vis team's concern that gaining buy-in for deployment from gatekeepers would be a major challenge; in contrast, the GER team was unsurprised.}

\subsubsection{R-Curate}

All exhibit requirements rely on displaying suitable content for the environment and its audience, including connecting to themes and other exhibits. As museum staff understand these aspects best, we want to provide them with the ability to curate the content displayed to visitors, even if they do not have technical expertise with computation. Our second requirement is to ensure that content creation, curation, and loading is easy, fast, and flexible (R-Curate).

\subsubsection{R-Engage}

Our third requirement is to engage visitors (R-Engage).
Visitor engagement is a prerequisite to all museum learning goals; before visitors can learn and enjoy, they have to engage. 
Although previous design studies that focus on the use of visualization in museums do situate their work in terms of targeting engagement and handling broad audiences, as discussed in \change{Section} \ref{sec:related-work}, the full implications of the shared dependencies from R-Engage to specific learning goals have not been sufficiently discussed in the visualization literature.

Engaging, for example through enjoyment or interest, gives museum visitors positive associations with science and learning, which our museum staff collaborators indicate is particularly important for younger visitors. \change{Such non-analytical purposes of visualization are important to consider \cite{baumer2022course}.} To engage with an exhibit, visitors must first notice it, understand that it is intended for them, and find it intriguing enough to investigate further.

Exhibit designers do not expect to achieve success with the intended learning goals of an exhibit with every visitor; exhibits where only a small percentage of visitors achieve an intended specific learning outcome beyond engagement are often considered highly successful. This success condition is a stark contrast with more formal learning environments like classrooms, where only a few students meeting intended learning goals would be a problematic outcome.

Similarly, visualization design studies with exploratory data analysis goals typically aim to support all target users for all goals and tasks. Our situation, where the dependencies between these goals lead to expected drop-offs at each step, is less common.

\subsubsection{R-Inspire}

Fostering positive relationships with science is a priority in museums, especially the family- and child-oriented ones. Our fourth requirement is to inspire curiosity in visitors through the presentation of a variety of interesting, and potentially surprising, pieces of information (R-Inspire). 
Even with on-the-spot learning, visitors may remember information later and investigate it further or share it with others. 
Museum staff are also interested in inspiring behaviour change and future careers, but studying these long-term impacts is outside the scope of this study.

\subsubsection{R-Compare}
\label{sec:tasks-compare}

Directly supporting the finalized primary task, to learn about deep time by comparing varied-magnitude time periods, is our final visitor-facing requirement (R-Compare). We deemed it feasible to design an exhibit to accommodate and promote this comparison, inspired by Resnick et al.'s classroom exercise \cite{resnickDNS2017}. This requirement specifically
addresses the targeted learning goal of this project, in contrast to the previous four extremely general requirements. 

\subsection{Data Abstraction}
\label{sec:data-abstraction}

The final requirement of supporting comparison of varied-magnitude time periods (R-Compare) is not tied to any specific dataset. Rather, the content curation requirement (R-Curate) reflects the need for museum to create appropriate datasets that tie in to their own museum's collection and inspire visitors (R-Inspire). We identified characteristics that curated data must have to address the Compare requirement. 

We define a dataset as a list of values, split into ranges. 
A \textbf{value} is a single \change{data point}, defined by its measure: a \change{single number}. In addition to a measure, each value also has a name. \change{A \textbf{range} is a pair of start and end measures. Ranges contain values whose measures are between the range's start and end measures.} A dataset is split into ranges by indicating which values delimit the ranges. The entire range can optionally be given a name. 

The DeLVE data abstraction requires the ranges of curated datasets to have three key characteristics: ranges must be \textbf{monotonic}, meaning they are ordered from smallest to largest \change{start values}; \textbf{contiguous}, meaning neighbouring ranges share edges; and \textbf{disjoint}, meaning individual ranges do not overlap. This abstraction is sufficiently general that it could apply to non-temporal scenarios as well as the motivating requirement of temporal comparison.

Moreover, we suggest that curators use at least a few ranges, ensure between 3-12 values in each range, and that the ranges are approximately divided according to powers of ten.

\section{Design}
\label{sec:design}

We document our design methods, present the Connected Multi-Tier Ranges idiom at the core of DeLVE, and discuss the design of the system as a whole. We then provide design rationale, and comment on the implementation and architecture. 
\SuppSystem~contains further discussion of DeLVE's design evolution, additional figures, and additional implementation details.

\subsection{Methods}

The design and development of DeLVE and its core idiom began after concluding the project preparation stage. We used paper-based prototyping for the first two weeks, then switched to rapid iteration via digital prototypes for the next 7 months. \change{During this initial development stage, we used a sample dataset created by the research team which evolved alongside the prototype and informed future dataset curation. Once museum staff had curated their own datasets, we used them for testing.} The design team met weekly throughout this period. We also procured additional feedback and suggestions through presentations to experts in visualization, HCI, and informal learning. 

After analyzing and reflecting on the deployment results at M-Bio, we did a design iteration just before deployment at M-Sci; the final iteration came after reflecting on what we learned after 5 more months of deployment. 

\begin{figure*}[t]
    \centerline{\includegraphics[width=.75\textwidth]{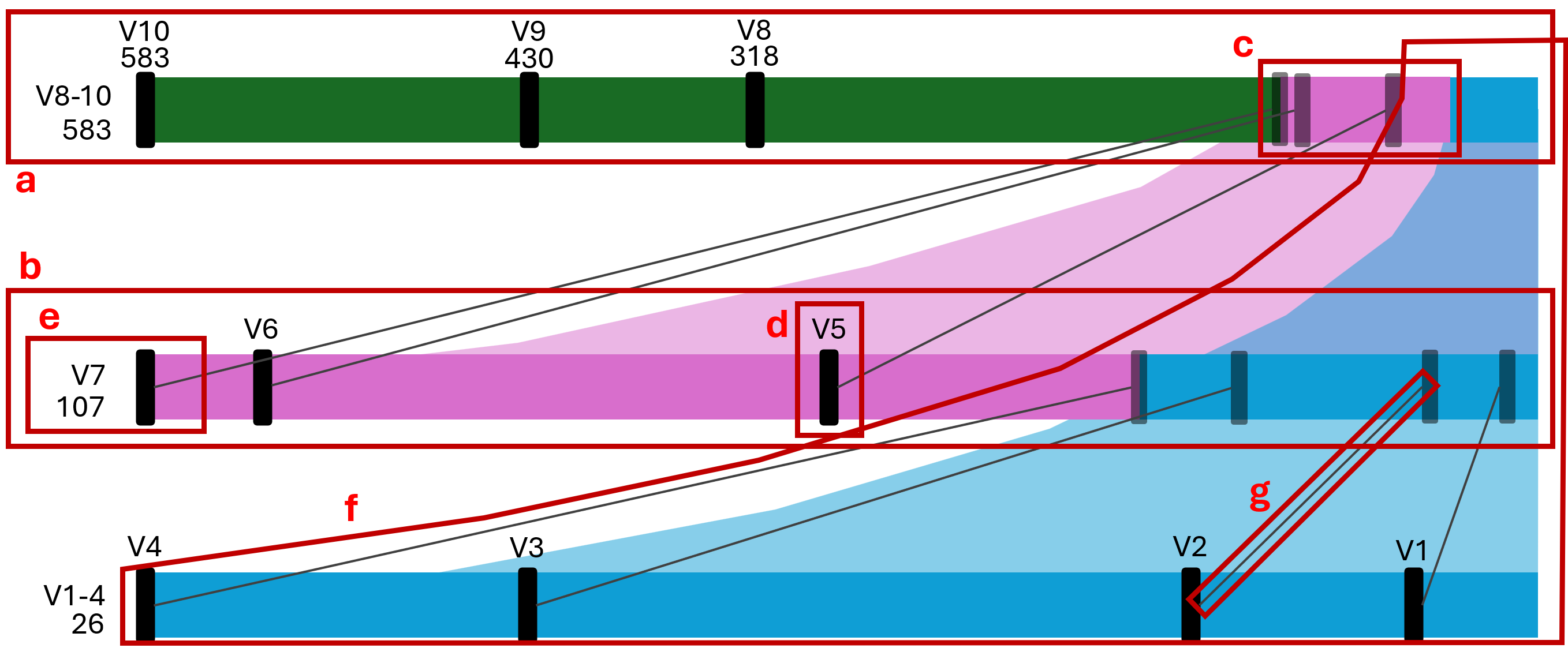}}
    \caption{\change{The Connected Multi-Tier Ranges (CMTR) idiom diagram with its components annotated. V1-10 represent values.} (a) The active tier. (b) An archived tier. (c) A segment. (d) A labelled value name marker. (e) A labelled name and measure marker in the tier title position. (f) A relation curve. (g) A relation line.}
    \label{fig:cmtr-diagram}
\end{figure*}

\subsection{Idiom: Connected Multi-Tier Ranges}
\label{sec:design:technique}

The core of the DeLVE interface is the \textbf{Connected Multi-Tier Ranges (CMTR)} idiom, a new technique that we propose for comparing varied-magnitude time periods (R-Compare). Although it was motivated by the specific need to showcase deep time, it could also be used for other scenarios where large scale-item ratios are in play~\cite{solen2024designspace}. 

Figure~\ref{fig:cmtr-diagram} shows \change{a diagram of the CMTR idiom}, with annotations highlighting the \delete{CMTR} components. The CMTR idiom consists of stacked tiers connected by relation curves and relation lines.
The topmost tier is the \textbf{active} tier, annotated as Figure~\ref{fig:cmtr-diagram}a, and below it can be multiple (zero or more) \textbf{archived} tiers. Figure~\ref{fig:cmtr-diagram} contains five of archived tiers, one of them annotated as Figure~\ref{fig:cmtr-diagram}b.
 
A \textbf{tier} is a multicoloured line with markers and labels. It \change{is composed of one or more ranges, each of which is encoded by a differently-coloured \textbf{segment}, as indicated in Figure~\ref{fig:cmtr-diagram}c. The measures in the dataset are encoded using linear horizontal position with larger values on the left, such that each segment spans between its left edge at the position of its range's start value and its right edge at the position of its range's end value. As such, the leftmost segment in a tier represents the range with the largest start measure, and a tier's start measure is the start measure of the leftmost range.}

Values are encoded with \textbf{markers}, namely rounded rectangular boxes. The values in a tier's \change{leftmost} range\change{, the range with the largest start measure,} are shown with opaque black markers, as indicated by Figure~\ref{fig:cmtr-diagram}d, with name labels just above them. The values in the second \change{leftmost} range are shown with translucent markers that have no labels. Values in the \change{other} ranges are not visually encoded in that tier. In the active tier, marker labels include the value measure itself below the name; these are times in the DeLVE use case. 

If the \change{leftmost} range in an archived tier has a title, that appears on the far left; otherwise the name and measure of \change{the range's leftmost} value will be shown there. 
Value measure labels here use words such as \textit{millions} or \textit{billions} rather than displaying numbers with many zeroes.

Each tier encodes all ranges of the tier below it plus the next contiguous range. Each range is \change{leftmost} in only one tier, and in the tier above that it is the second \change{leftmost}. Tiers are connected by colour-filled relation curves, of which there is one for each range. A \textbf{relation curve}, shown in Figure~\ref{fig:cmtr-diagram}f, connects the start points of its range's segments across all tiers that include that range, with the region below the curve filled with the range's colour, as with a filled area chart. 

Each value is encoded in two tiers. In the tier where the value's range is \change{leftmost}, the markers are opaque black, and in the tier where the value's range is second \change{leftmost}, they are translucent. In addition to relation curves, tiers are connected by grey \textbf{relation lines}, as indicated in Figure~\ref{fig:cmtr-diagram}g. Each value has one relation line, which connects the value's associated opaque black marker in a tier above to its translucent marker in the tier below.

The active tier is a special case as it has no tier above it. The \change{range with the largest start measure} across all tiers only appears here, so there is no relation curve for it. Also, values from the largest range across all tiers only appear here, so there are no relation lines rising above them.

When the user interacts with the system to advance it, the next value appears on the active tier through an animated transition. 
The \change{marker for the next value is added at the far left and the} tier rescales to include it. Simultaneously, all existing segments resize, the segment for \change{leftmost} range grows to the right, the segments for the other ranges become shorter and shift to the right, existing markers move, and a new labelled marker fades in.
Figures~\ref{fig:delve-walkthrough}a-b show the start and end of such a transition. 

A dynamic animation triggers when the newly added value is outside the currently displayed range in the leftmost segment of the active tier, with the following changes happening simultaneously. A new segment, encoding the new \delete{largest} range, grows within the active tier, starting from the leftmost point and growing towards the right. As it grows, the entire tier rescales \change{to accommodate the new segment}. The new value appears in this new segment. The value markers in the active tier's now-second largest segment fade to translucent with the labels fading out completely. A copy of the previous version of the active tier instantiates behind the existing active tier and gradually animates downwards to its destination below the active tier and above all other archive tiers. A new relation curve gradually appears as the new archive tier moves downwards, stretching between the \change{leftmost} segment in the new archive tier and the now-second \change{leftmost} segment in the active tier. New relation lines also gradually appear by stretching between the event markers in those two segments. Three key frames from this animation are shown in Figures~\ref{fig:delve-walkthrough}d-f.

Tiers are vertically stacked with \change{the tier with the largest start measure} on top and the \change{tier with the smallest start measure} on the bottom. The spacing between the tiers is proportional to the multiplicative difference between the \change{start measures} of the tiers. When archived, previous tiers move down in screen space. 

The CMTR idiom combines visual encoding and interaction to allow users to gradually step through the list of values until the entire dataset is simultaneously visible after reaching the final value\delete{, as shown in Figures~\ref{fig:delve-overview-teaser} and \ref{fig:cmtr-diagram}}.

\subsection{DeLVE Exhibit}

We present DeLVE, the Deep-time Literacy Visualization Exhibit, in Figure~\ref{fig:delve-overview-teaser}. From top to bottom, it contains a title and optional subtitle, an overall separate timeline, an instantiation of the Connected Multi-Tier Ranges (CMTR) idiom, a media box containing additional description and imagery for the active event, and buttons for navigation. The CMTR idiom, as discussed in Section~\ref{sec:design:technique}, uses the majority of the screen space. As DeLVE focuses on deep time, our ranges are time periods and our values are events. \change{Additionally, we reduce the width of the archive tiers to make space for the media box and buttons.}

\begin{figure*}[t] 
    \begin{subfigure}{0.33\textwidth}
        \includegraphics[width=\linewidth]{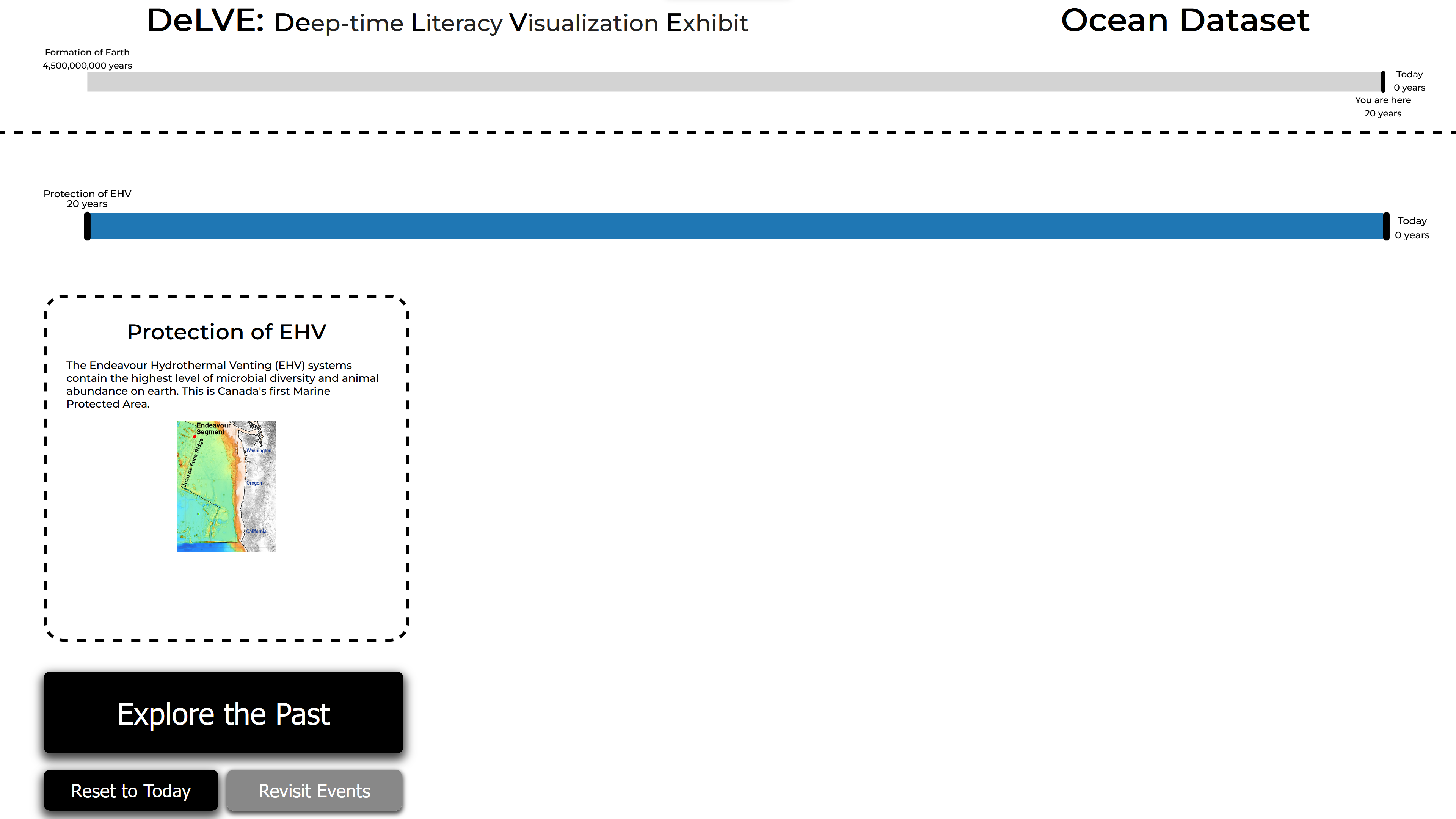} 
        \caption{}
    \end{subfigure}
    \begin{subfigure}{0.33\textwidth}
        \includegraphics[width=\linewidth]{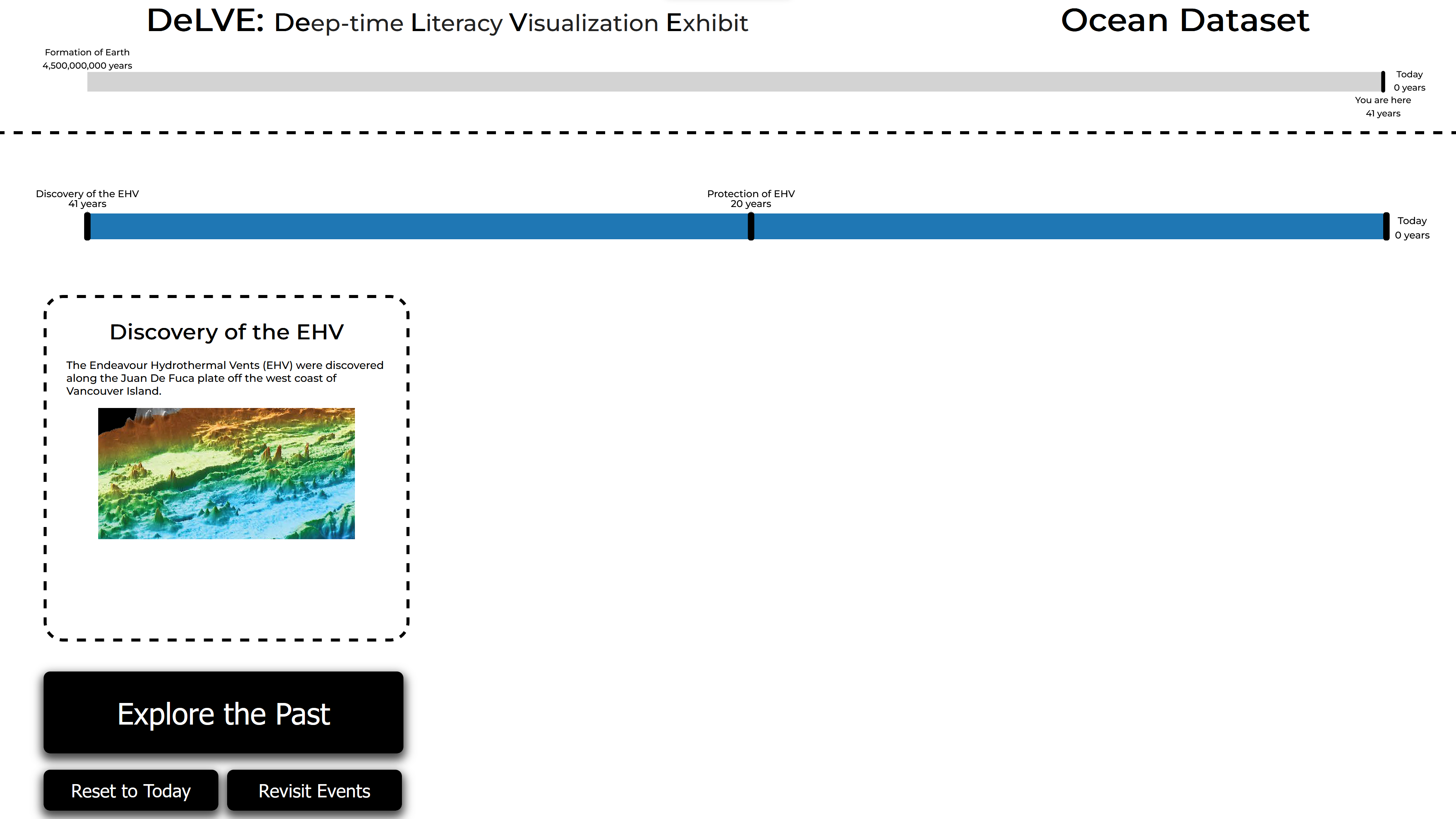}
        \caption{}
    \end{subfigure}
    \begin{subfigure}{0.33\textwidth}
        \includegraphics[width=\linewidth]{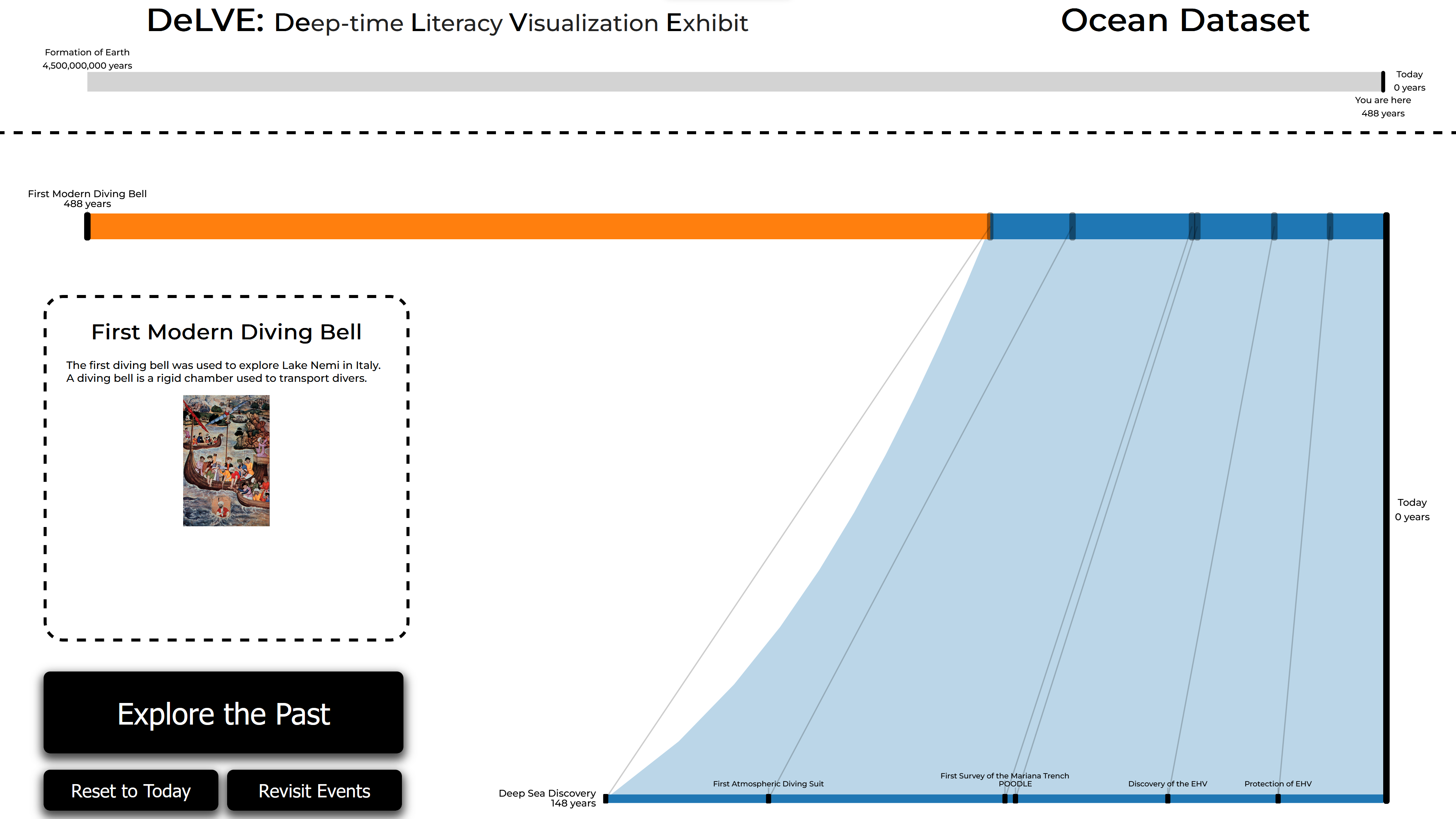} 
        \caption{}
    \end{subfigure}
    \begin{subfigure}{0.33\textwidth}
        \includegraphics[width=\linewidth]{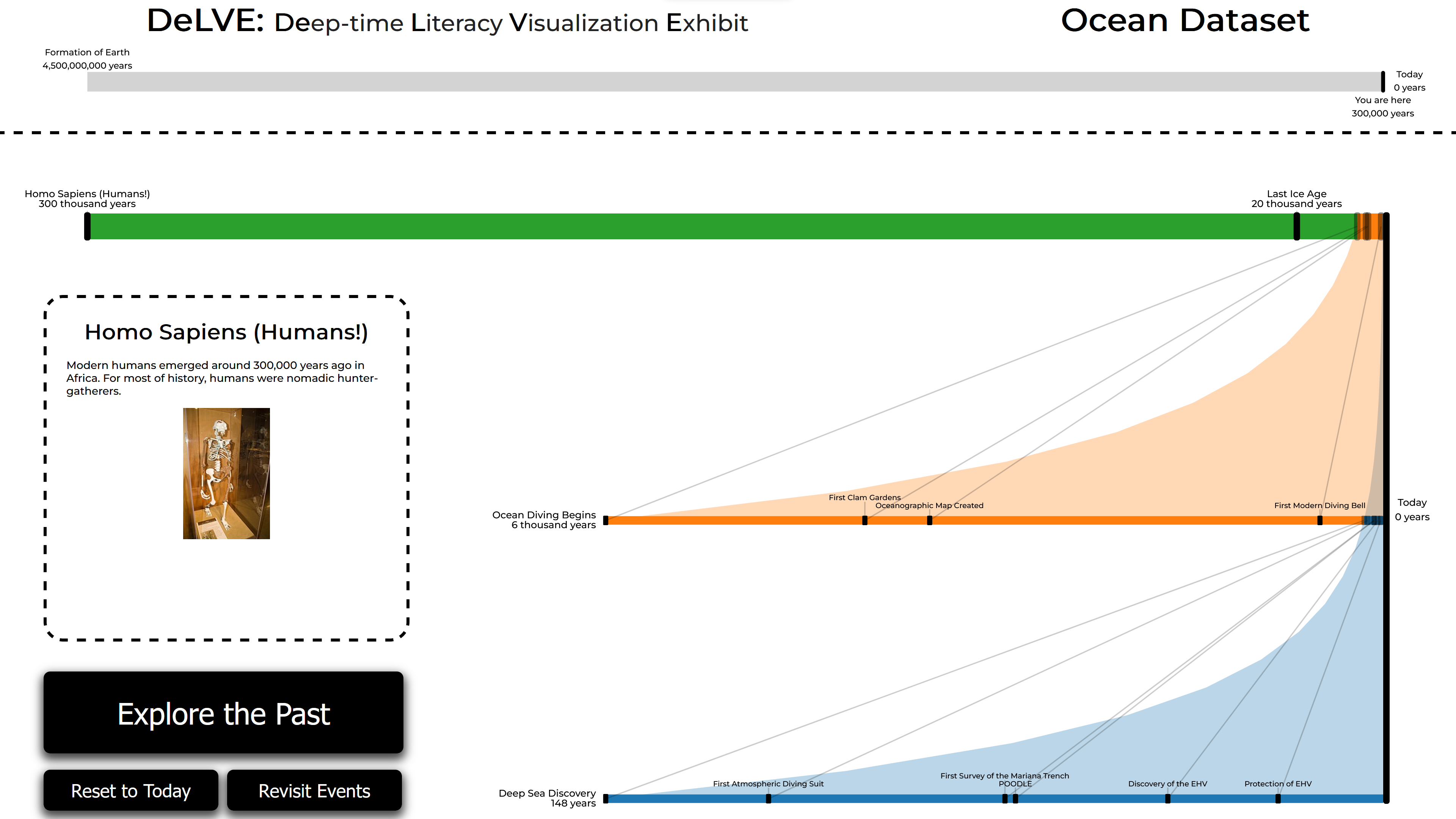}
        \caption{}
    \end{subfigure}
    \begin{subfigure}{0.33\textwidth}
        \includegraphics[width=\linewidth]{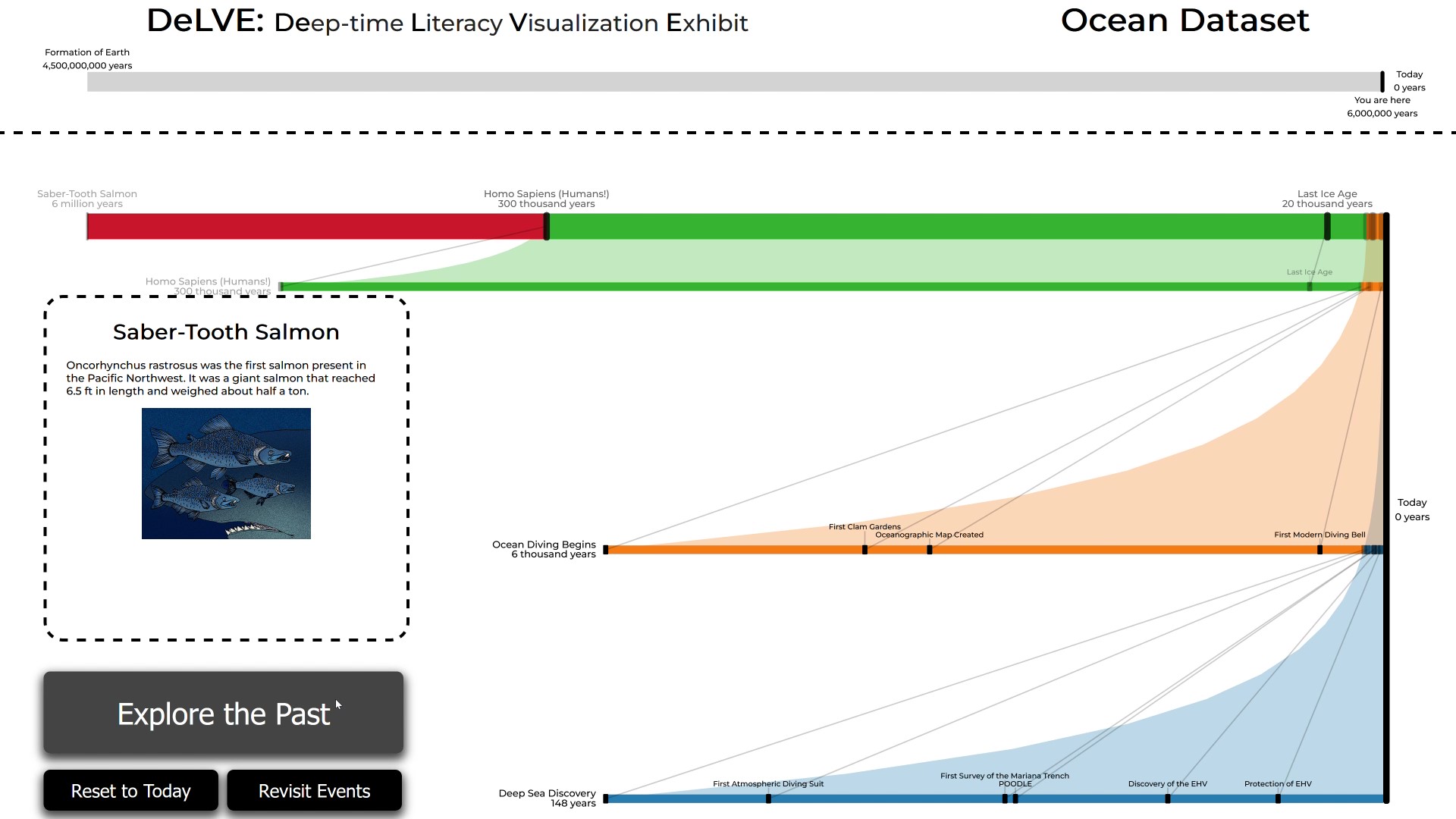} 
        \caption{}
    \end{subfigure}
    \begin{subfigure}{0.33\textwidth}
        \includegraphics[width=\linewidth]{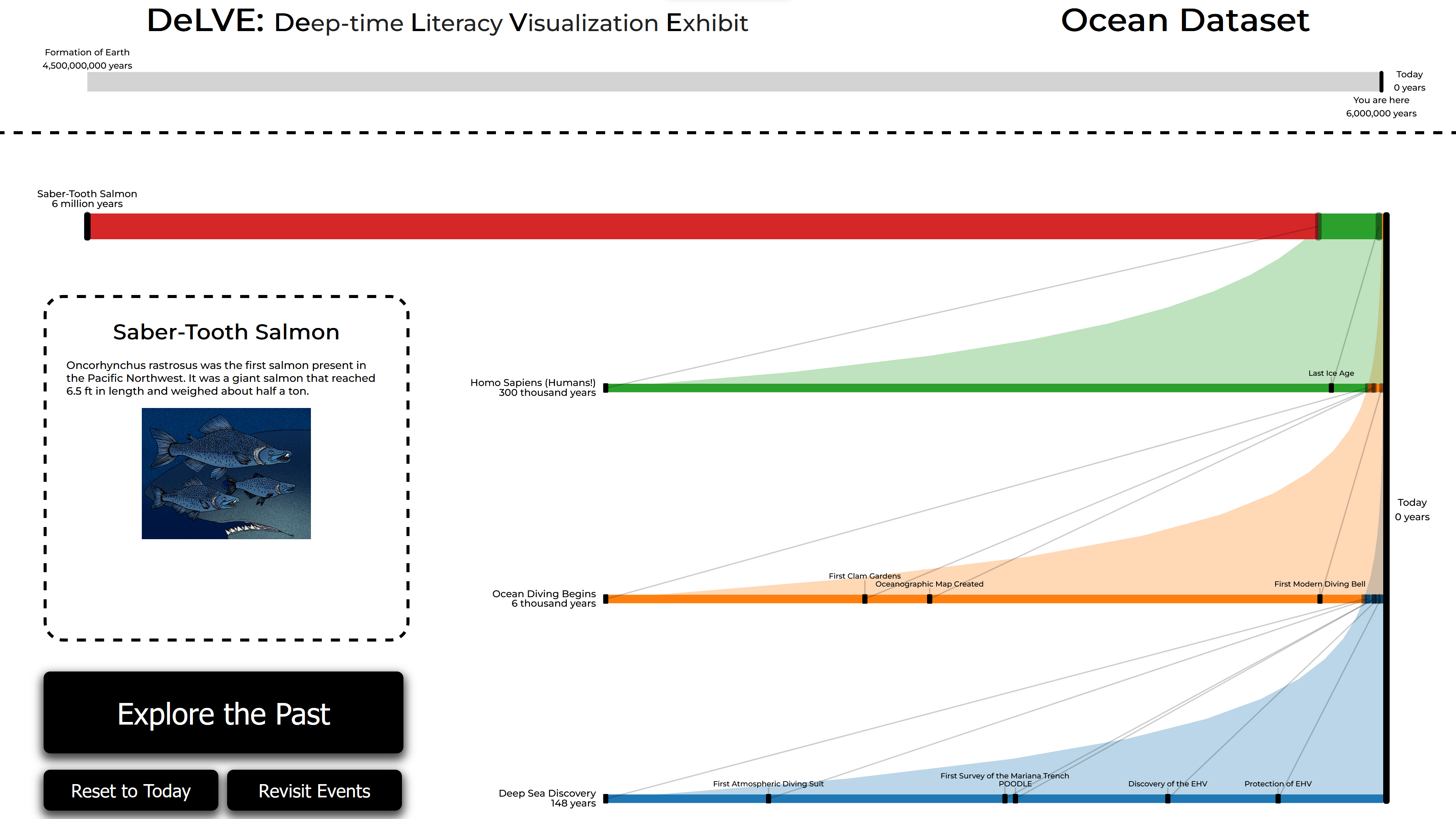}
        \caption{}
    \end{subfigure}
    \caption{Key frames from a walkthrough of DeLVE. a) The initial state, with only a single event. b) After clicking \textit{Explore the Past} once. c) After progressing to the second tier. d/e/f) Three frames from a single animation showing animation progressing from the most ancient event in the third tier to the most recent event in the fourth tier.}
    \label{fig:delve-walkthrough}
\end{figure*}

The \textbf{media box} shows event details for the active event. It contains the event name, description, and image. The details can be curated (R-Curate) to inspire visitors (R-Inspire) and connect with specific topics.

The largest button, labelled \textit{Explore the Past}, progresses the entire system to the next event in the dataset. It updates the active event and triggers all relevant dynamic animations. The \textit{Reset to Today} button removes all archive tiers and all active tier events except for the one closest to present day, rescaling the active tier to match. The \textit{Revisit Events} button changes the active event to the one that appeared directly before it. A revisited event is visually highlighted with a hollow rounded black box around its labels. Multiple revisits only cause changes in which event is highlighted, not with inverse animations to ``roll back'' time. After revisiting, when the \textit{Explore the Past} button is used again, the highlighting happens in reverse until the user reaches a new event and animations resume. We carefully chose the wordings of these buttons to indicate to users that pressing, for example, \textit{Explore the Past} will show an event further back in history rather than one that they had previously seen.

In addition to the buttons, pressing on an event label or marker revisits directly to that event. Pressing anywhere else on the screen, where there are no event labels and markers or buttons, will trigger the \textit{Explore the Past} button to be pressed, including a visually pressing the button as if the user had interacted with it directly. 

The overall separate timeline is a multicoloured line with three labels, shown \change{at the top of} Figure~\ref{fig:delve-overview-teaser}. Similarly to the CMTR tiers, it uses the same coloured segments to show time periods in the same horizontal order on a linear scale. It also shows the highlighted event with a labelled marker, again a black rounded rectangular box. Unlike tiers, the time range is static and covers all events. The most ancient event in the dataset is always visible on the far left of this timeline and present day, labelled \textit{Today}, is always visible on the far right of it. The active event is labelled with \textit{You are here} rather than the event's name. Segment colours are only visible to the right of the time of the CMTR view’s most ancient displayed event across all tiers. 

DeLVE supports three modes: interactive, animated, and dynamic. In interactive mode, progression through events is controlled solely by the user pressing buttons. In animated mode, progression through events is controlled fully automatically at regular intervals. In dynamic mode, progression through events is controlled by the user pressing buttons, unless there is no interaction for a configurable amount of time, at which point it begins automatically progressing. Automated progression in dynamic mode can be stopped by the user at any point by pressing the \textit{Let Me Interact!} button, which is intended to encourage interaction.

Figures~\ref{fig:delve-walkthrough}a-f show six key frames from a walkthrough of DeLVE. The video included in supplemental materials shows the look and feel of DeLVE.

\subsection{Design Rationale}

DeLVE's visitor-facing interface addresses the three visitor-facing requirements: R-Engage, R-Inspire, and R-Compare. We considered both initial engagement and prolonged engagement when thinking about R-Engage. We intend for three aspects of design to support initial engagement. We use animation and colour \cite{maMF2019} to catch visitor attention and images in the media box to gain initial interest. Once visitors have approached the exhibit, we rely further on the images and animation to keep their attention, as well as the events themselves which we intend to be familiar and interesting. The media box is the primary facilitator for R-Inspire, as it can present information and diagrams to the visitor. Absolute and relative timing of events is the secondary facilitator of this requirement, as visitors may find them interesting or surprising.

Finally, we designed both the CMTR idiom and the entire DeLVE system to emphasize comparison between scales and events in support of R-Compare. Users can compare the lengths of the segments with shared colours across tiers, an instantiation of \textbf{unitization} from concrete scales, or the vertical spacing between pairs of tiers, an instantiation of \textbf{analogy} from concrete scales \cite{chevalierVG2013}. They can follow the logarithmic shape of the relation curves to see the relative length of a time period on exponentially increasing scales, and can study the angles of the relation lines that connect the tiers. In addition, the overall separate timeline shows exponentially increasing change on a linear scale. When the active event changes, the event marker moves to its new spot. However, it does not visibly move until the active event is in a very ancient time period, as all changes between events that are multiplicatively close to present day subtend less than one pixel. We intend for this view to cause surprise in users at how long it takes to see visible progress on the linear scale of Earth's history.

We note that geologist norms would use the opposite placement pattern, where the longest time range is typically on the bottom, building up to the shortest one on the top. We chose this placement so that viewer attention can stay near the top of the screen to align with typical screen fixation patterns, both because we believe the general public is unaware of the geologist convention and to mirror Resnick et al.'s teaching exercise \cite{resnickDNS2017}.

\subsection{Implementation and Architecture}

The front end of DeLVE was created with JavaScript, HTML, and CSS, primarily using the D3 library to create the visualizations. The interface we described in the previous section covers the visitor-facing side of the front-end. Museum staff and other administrators will also use the welcome and settings pages. The welcome page is the first page shown upon navigating to the URL, and users can choose to see a sample dataset in DeLVE without configuring any settings or to use a custom dataset and configure custom settings. Using Google Sheets and a simple interface for data uploading allows for museum staff to curate their datasets, meeting R-Curate. If curators do not provide any range delimiters, DeLVE will automatically compute them to roughly group values according to powers of ten. 

We store logs of usage of DeLVE on a server hosted on the vis team's university department servers as a remote-only Linux virtual machine. The server code is in JavaScript using Node.js and Express. 

\begin{figure*}[t] 
    \centering
    \begin{subfigure}{0.24\textwidth}
        \includegraphics[width=\linewidth]{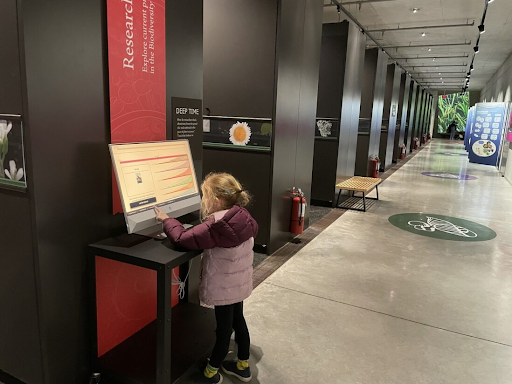} 
        \caption{}
        \label{fig:deploy-bio}
    \end{subfigure}
    \begin{subfigure}{0.24\textwidth}
        \includegraphics[width=\linewidth]{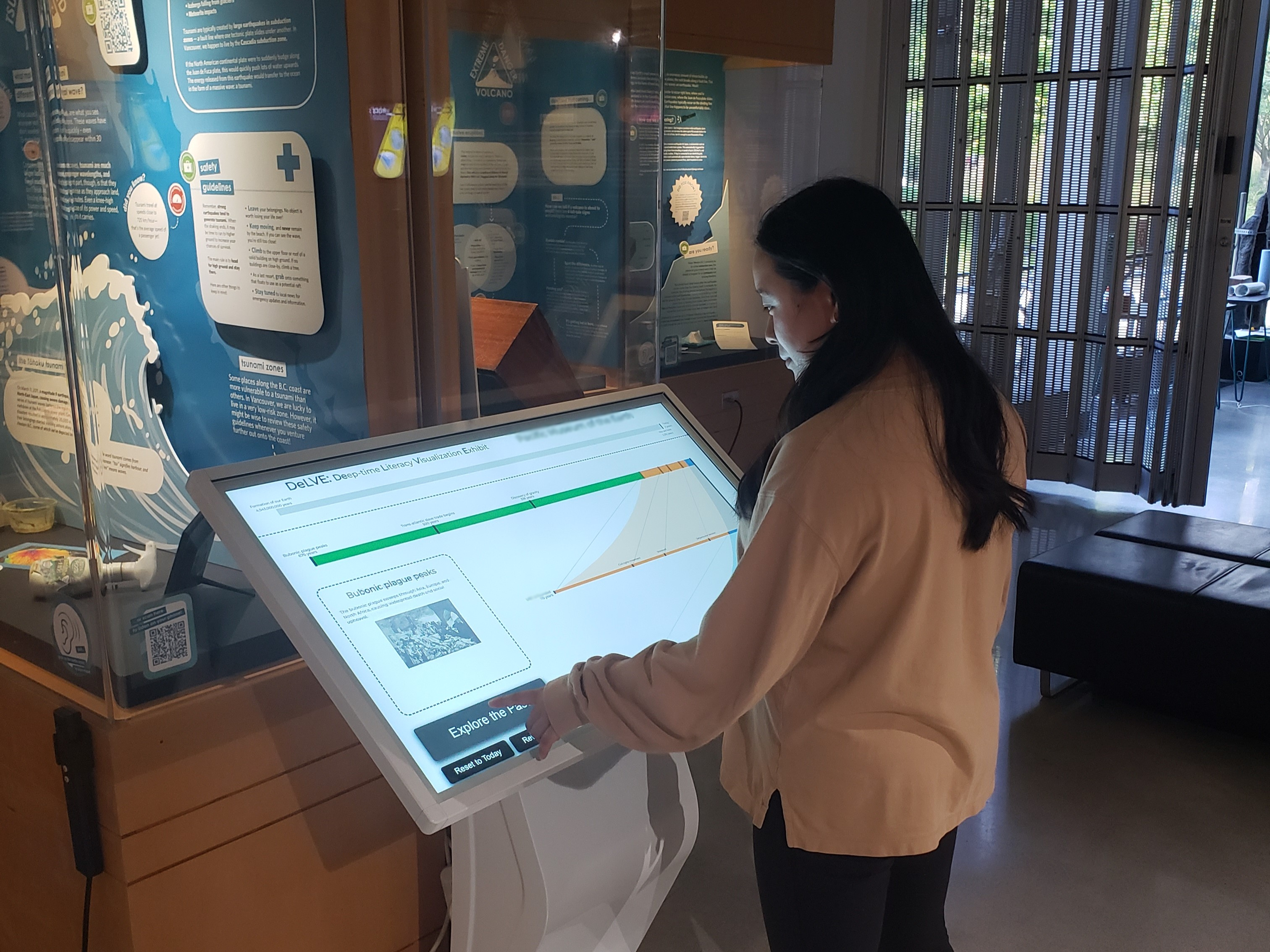}
        \caption{}
        \label{fig:deploy-geo}
    \end{subfigure}
    \begin{subfigure}{0.24\textwidth}
        \includegraphics[width=\linewidth]{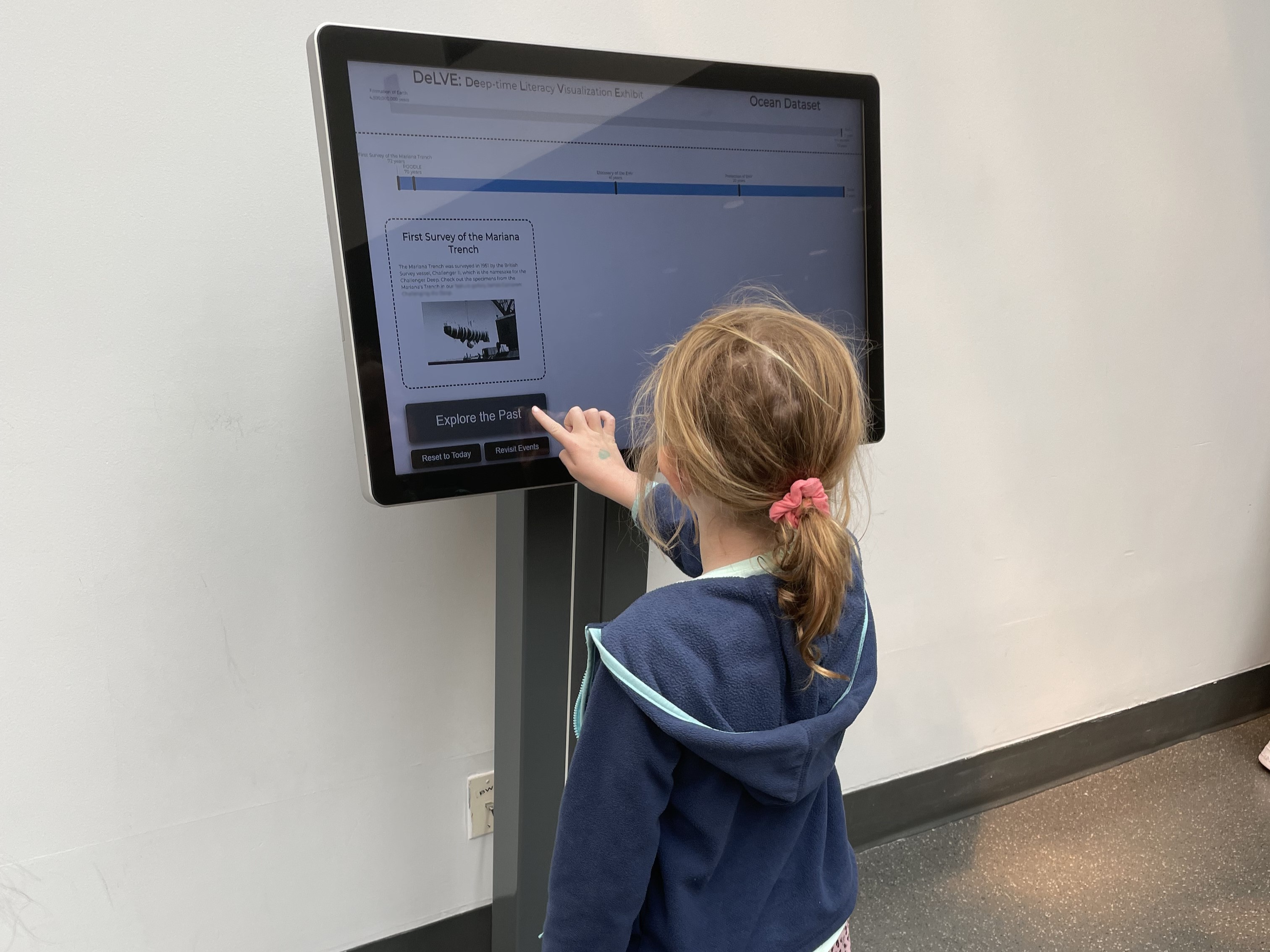}
        \caption{}
        \label{fig:deploy-sci}
    \end{subfigure}
    \begin{subfigure}{0.24\textwidth}
        \includegraphics[width=\linewidth]{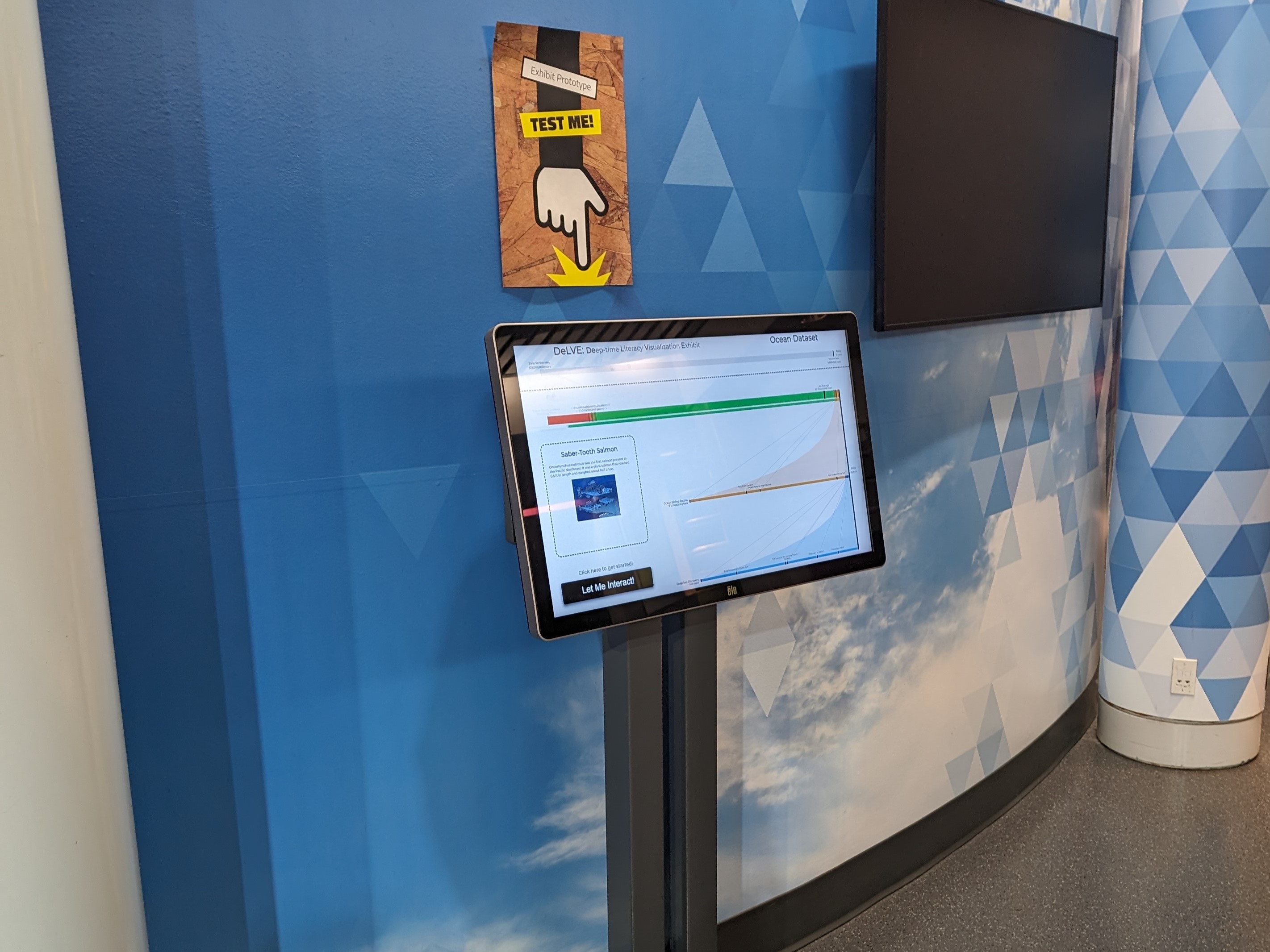}
        \caption{}
        \label{fig:deploy-sci-2}
    \end{subfigure}
    \caption{DeLVE, deployed as a digital kiosk in (a) M-Bio, (b) M-Geo, and (c)/(d) M-Sci.}
    \label{fig:deploy}
\end{figure*}

\section{Deployment}
\label{sec:deployment}

After completion of the initial DeLVE prototype in February of 2023, we met with our museum collaborators and made deployment plans. See \SuppDeployments~for additional details on DeLVE's deployment.

\subsection{Deployment Process}

\change{Through the expert interviews, we found that museum staff at all three museums were very enthusiastic about collaboration. These domain experts immediately offered to devote further time to talking with us in support of the project, and advised us that there were unlikely to be barriers to eventual deployment so long as we communicated sufficiently. This outcome stood in contrast to the vis team's concern that gaining buy-in for deployment from gatekeepers would be a major challenge; in contrast, the GER team was unsurprised as their past collaborations had led to an awareness of the museum priorities and a reciprocal approach where both parties were looking to benefit each other.}

We began the deployment process by holding deployment approval meetings with museum staff; after followup communication, these led to deployment approvals at all three museums: a biology museum (M-Bio), a geology museum (M-Geo), and a science centre (M-Sci). \change{Receiving approval to deploy in all three museums shows success in R-Deploy.} While museums and science centers differ \cite{ynnermanLB2020}, below we refer to them all as museums for simplicity.

Staff at all three museums were enthusiastic about our design \change{due to our extensive background work and their general interest in supporting research efforts,} and wanted to deploy an instance of it in their institutions. 
See \SuppMethods~ for the details and outlines of the deployment approval meeting presentations and the guiding questions for the discussion, as well as supplemental material for the transcript files of those discussions.

Staff at all three museums contributed time and resources to deployment. \delete{Staff at M-Bio and M-Sci provided kiosk hardware to deploy DeLVE on. The design team purchased kiosk hardware for DeLVE's deployment in M-Geo.}
Museum staff also devoted time to dataset curation, including collective decisions on reading level, text description length, and number of events as well as curation of actual data\change{, fulfilling R-Curate}. We also encouraged museum staff to incorporate recent events, which happened within visitor lifetimes, into their datasets to increase the personal relevance of the exhibit \cite{meier2017individual}.

Staff at M-Bio and M-Sci have committed to long-term deployment of DeLVE on their own hardware\change{, showing additional success in R-Deploy}. To support this deployment, they are discussing final deployment locations within the museums. In addition, staff at M-Sci are developing a new dataset to fit with the context of the new location. They are also designing signage to place around DeLVE to bring attention to it and provide further information.

\subsection{Museum Contexts}

We deployed DeLVE in all three museums on large touch screen kiosks, although the specific hardware differed. See Figure~\ref{fig:deploy} for images of the kiosks in context. \change{Museum staff made deployment location decisions based on available space, proximity to other relevant exhibits, and availability of utilities such as power and internet.} The three museums had important differences. 

M-Bio consists primarily of rows of cases and drawers with biological specimens. General audiences are welcome, and they provide guided school-group tours and themed events.

M-Sci consists of differently-themed rooms with science-focused educational activities and exhibits. They provide visitors with games and other experiences to encourage them to think about high-level ideas such as their connection with their community or the environment. M-Sci is more family-oriented, so their audience includes more younger children than the other museums. The DeLVE kiosk in M-Sci initially stood in a hallway between exhibit galleries, near the entrance to a gallery on deep ocean exploration (M-Sci-hallway context). \delete{Most visitors in this area move through that space without stopping; most of those who stop focus on resting rather than new activities.} Later, M-Sci staff moved the kiosk to a different room themed around optical illusions and physical puzzles (M-Sci-puzzles context). \delete{While the other exhibits in this room were disconnected from the content of DeLVE's M-Sci dataset, most visitors in the area were actively looking for activities to engage with.}

M-Geo does not have a self-contained space; it is embedded into rooms and walkways across two university buildings \change{with most who enter the space simply passing through. Its} exhibits consisting mostly of geological objects and geology-related text and images. \delete{The space is primarily used by university students, staff, and faculty.} The museum provides guided tours and supports some coursework activities. The vast majority of people who pass through the space are on their way elsewhere, and do not engage with the exhibits; we eventually focused our attention more on the other two museums where we could gather more useful data.

At M-Bio, the other exhibits are heavily text based and incorporate technical terms, and thus are more similar to DeLVE than those at M-Sci, where exhibits are more playful and use simpler language. The museum spaces also have architectural differences. In M-Bio, the space is mostly made up of long hallways painted dark colours with little natural light. M-Sci is bright, colourful, and open, with many skylights.
There are multiple differences between the audience distributions and behaviours of visitors between M-Bio and M-Sci, including the age distribution of visitors which skews younger for M-Sci.
An additional difference in museum curation between M-Sci and M-Bio is complexity level consistency. Exhibits at M-Bio are mostly consistent with each other in style and complexity level. At M-Sci, different rooms and different exhibits appear to be designed for different age groups, so the range of visitors can seek out areas appropriate for them.

\section{Evaluation}
\label{sec:evaluation}

We evaluated DeLVE using data from visitor observations and system logs, collected during DeLVE's deployment. We first discuss the methods of this data collection and analysis and then their results, then present our analysis for evidence of meeting our visitor-facing requirements and for differences between deployments.

\subsection{Methods}

We conducted our observational study in M-Bio from April 2023 to February 2024, in M-Sci from August 2023 to March 2024, and in M-Geo in December 2023. We observed visitors interact or not by their own choice and did not intervene.
During observation sessions, a researcher sat nearby DeLVE and recorded any observations of individuals or groups who came within three meters of the exhibit. The researcher tallied participants who came close to DeLVE but did not engage with the exhibits and took more detailed notes on those who did engage \cite{sultana2023preliminary}.

We conducted 25 observation sessions, totalling nearly 37 hours of observation, and resulting in 95 observations of visitors engaging with DeLVE without intervention. We conducted 16 of the sessions, making up over 24 hours and 44 observations, at M-Bio; 3 of the sessions, making up just under 5 hours and 13 observations, at M-Sci's first deployment location (hallway); 5 of the sessions, making up just under 6 hours and 38 observations, at M-Sci's second deployment location (puzzles); and 1 of the sessions, making up 1.5 hours and 0 observations, at M-Geo. Transcribed versions of the observations are available in the supplemental material, and full forms and protocols are in \SuppMethods. 

We calculated statistics using observation tally counts. \change{The first author} coded our observations with respect to our three visitor-facing requirements. 
We also consider statistics from our system trace logs, which we group into interaction sessions delimited primarily by breaks in logs. We use both observation and trace log data in both of our analyses, requirement-focused and difference-focused.

\subsection{Requirement-Focused Analysis}

We now report findings from our evaluations in terms of our three visitor-facing requirements, discussing both those that led to modifications to the design and overall findings. 

\subsubsection{R-Engage}

Engagement is a prerequisite for learning and enjoying, and both previous museum visualization work and our museum staff collaborators note it as a challenge. We now discuss our overall findings from all three museum deployments, including 95 observations of visitor interaction across M-Bio and M-Sci, in terms of engagement.

Many observed visitors did not appear to notice DeLVE at all. Of the smaller number who visibly noticed DeLVE, many would look away immediately or only pause momentarily before moving on. Of those who noticed DeLVE, about 18\% chose to engage with DeLVE, either by an extended watching of the animations, directly interacting using the buttons, or a combination of both. Our museum staff collaborators at M-Bio confirmed that our engagement levels were on par with other similar exhibits, and their and M-Sci's commitment to long-term deployments of DeLVE show that staff at the institutions see the exhibit as successful.

Among those who engaged with DeLVE, 75\% spent around 30 seconds or more with it. Thus, the majority of engaged users would be considered ``hooked'' \cite{horneckerS2006}, meaning museum-visitor engagement beyond a ten-second threshold. 25\% of these participants spent two minutes or more engaging with DeLVE, with 3\% spending over five minutes. The mean length of the estimated interactions from the logs is 94 seconds and the median is 25 seconds.

We made minor iterations on DeLVE's design to increase engagement based on our observations. We made the buttons more visually salient to make DeLVE's interactivity more obvious, decreased the timeout before initiating the automatic animation mode to make DeLVE more likely to be animating and catch participants' eyes, and implemented responsivity to touches anywhere on the screen to engage participants who did not initially interact with the buttons. See \SuppSystemEvol~for further detail on DeLVE's evolution. 

The long interaction times indicate success with engagement and that accomplishing further goals is feasible.

\subsubsection{R-Inspire}

We observed clear indications of curiosity among visitors who engaged with DeLVE, at two levels. First was the initial curiosity caused by the colours and animation, which is DeLVE's entry point \cite{blockHPDES2012}. Once visitors began interacting, 24\% of participant groups showed signs of curiosity about the actual content within DeLVE, which was our goal. Observed curiosity took many forms, including that of facial expressions of surprise such as raised eyebrows and open mouths and behaviours indicating enjoyment such as laughing. Other participants indicated curiosity by their chosen topics of discussion with other group members, either by asking each other questions, mentioning specific information they found interesting, or educating each other. We also observed participants using their phones to search the internet or take pictures, potentially to investigate something further after they ended their interaction. We found it was more than three times more likely for a group to show behaviours indicating curiosity than for an individual to do so. However, most groups had two or more members, making up 54\% of all observations. Given that many of the indications of curiosity that we noted in our observations involved communication between multiple individuals in a participant group, we believe that many of the lone individuals we observed may have become curious while interacting with DeLVE but did not express this curiosity due to a lack of other group members to express it to.

These behaviours indicate that DeLVE successfully inspired participants' curiosity about the content in DeLVE's datasets.

\subsubsection{R-Compare}

We observed 7 participant groups, or about 7\% of all participant groups, directly comparing scales, either verbally or via gestures. Visitors talked about individual ages and times and gestured the timelines on the CMTR and on the overall separate timeline, sometimes appearing surprised by the ages they saw or saying so out loud to another visitor. Similar to R-Inspire, our indications of comparison relied on visitors having someone to communicate with, so individuals may have made comparisons without externalizing them. Of the groups with more than one individual, our observations of comparison make up 13\%. Further, many groups, despite not engaging alone, did not \change{visibly or audibly} communicate throughout their engagement. Of those that did communicate \change{verbally}, many groups discussed events without explicitly making comparisons, or their discussion was too quiet too hear or in a language other than English. Many others may have made comparisons without communicating them \change{in a way that was observable by the researcher}, even those in groups.

These observations of visitors show that DeLVE is successful in facilitating comparison between different time scales.

\subsection{Difference-Focused Analysis}

We \change{only analyze differences} between M-Bio, where DeLVE stood in a space with other exhibits with similar content; M-Sci's first deployment location, where DeLVE stood in a hallway far from other exhibits (M-Sci-hallway) and M-Sci's second deployment location, where DeLVE stood near other exhibits that were themed around puzzles and illusions and did not have similar content (M-Sci-puzzle). We did not observe any participant interactions in M-Geo, so we do not discuss it further here. See \SuppResultFigs~for figures showing observation breakdowns and system log statistics.

One major difference between our observations of participants at M-Bio, M-Sci-hallway, and M-Sci-puzzle was the number of individuals who noticed DeLVE, with far fewer at M-Sci-hallway. In contrast, M-Bio and M-Sci-puzzle had similar numbers of participants who noticed DeLVE. Similarly, we observed that most participants at M-Sci-hallway who engaged with DeLVE did so for less than a minute, in contrast to our observations of M-Sci-puzzle and of M-Bio to an event greater extent, where participants were much more likely to engage for one minute or more. We found similar results when analyzing for the amount of the dataset that participants clicked through and the number of event descriptions that participants read.

Analyzing durations from the log data, we again find that M-Bio durations are much longer, with a mean of 94 seconds, than M-Sci-hallway and M-Sci-puzzle durations, which have means of 42 and 37 seconds respectively.
We also note that the number of interactions per month in M-Sci-puzzle is higher than that in M-Sci-hallway despite observing over 60\% more visitors passing through the latter's proximity zone. Interestingly, we find that M-Sci-puzzles's median interaction duration in the trace logs is much closer to M-Sci-hallway's, and is in fact 15\% \textit{shorter}. This difference between our observations and trace logs may come from a type of interaction we observed very often in M-Sci-hallway but rarely in other deployments where participants would tap on the exhibit buttons as they moved past it. If multiple participants tapped on the screen within one minute of each other, we would count these button presses as the same session because, in the system trace log data, it would be indistinguishable from a single participant who tapped on a button twice.  

The low level of engagement at M-Sci-hallway is likely due to its location in a hallway. Because there are no other exhibits nearby, we observed most participants in the area simply passing through, focused on finding another room with exhibits to interact with in it. While M-Sci-puzzle's engagement levels were closer to M-Bio's than M-Sci-hallway's, they were still lower. This situation could be due to the difference in visitor age distribution: participants in M-Bio tended to be older. On one specific day in M-Bio, the largest audience in the museum was children on a school field trip. On that day, the distribution of observed behaviours was much more similar to that we observed in M-Sci-hallway than the other days in M-Bio. Another potential cause of this difference in engagement between M-Bio and M-Sci-puzzle is the latter's difference in content theme from other exhibits in the surrounding area. It is possible that visitors at M-Sci-puzzle who engaged with DeLVE ended their engagement early because of its lack of connection to the exhibits the visitors had recently interacted with.

\section{Discussion}
\label{sec:lessons-learned}

We now reflect on the design study, discussing the generalizable findings of our visitor studies and the differences between our design study process and previously documented approaches.

\subsection{Context Matters Immensely}

We found significant differences in visitor usage of DeLVE in different museum contexts. While previous museum visualization papers are informative, their design may have been received very differently and they may have produced different conclusions had they deployed the same exhibit in a different institution or even a different room within the same institution. Our findings echo the results of O'Reilly and Inkpen, where busy environments full of distractions yield different results than the focused attention possible in ``white rooms'' \cite{reillyI2007}. 

Museum exhibit designers need to understand the audience and context of the museum spaces they are deploying in to make an effective design. It is well known in museum visualization literature that museum audiences are very diverse, a finding reflected in our observations as well, but museum visualization designers \change{may still need to} consider the specific distribution of age, expertise, motivation, and other traits among that audience in order to tailor the design accordingly. The physical context of an exhibit \change{may also be a factor in} its success, and designers \change{should} consider what the visitors' expectations are within the exhibit space in order to create an effective and engaging exhibit, considering what types of exhibits are nearby, what kind of content is in those exhibits, and how the space is designed. Museum staff are experts in the audiences and contexts of their own museums, and can often inform us about them if we ask the right questions. Going forward, museum visualization designers should not be content with simply knowing their user pool is diverse and that they are deploying in a museum, but should investigate the details.

\subsection{\change{Concept-First} Design Study Methodology}

While conducting this design study, we observed that many aspects of our process differed from that described in the DSM of Sedlmair et al.~\cite{sedlmairMM2012}. Although we noticed these differences early on, after reflection we decided to continue the project without forcing ourselves to conform to the standard. We believe that these differences arose not because of an unsuccessful project, but because \change{our design study was triggered by the discovery of a concept to communicate, which is later refined into a presentation-focused task, and connection with collaborators and the acquisition of data happen later. In contrast, the DSM suggests a collaborator-first approach where finding collaborators occurs early, followed by the construction of data and task abstractions before beginning the Design stage. Oppermann et al.'s proposed data-first design study methodology starts with acquiring data early, with task abstraction and connection to collaborators coming afterwards, again before the Design stage \cite{oppermann2020data}.} 

\change{The concept-first design study applies primarily to presentation-focused design studies, as they involve the presentation of already-gleaned information, or concepts, rather than the gleaning of new information. However, presentation-focused design studies may also be triggered by connection with collaborators or the acquisition of data. The discovery of the concept to present is the defining trait of our process.}

We now discuss the \change{individual} stages in which our process differed from the DSM, providing alternative methodological guidance for presentation-focused design studies.

\change{\textbf{Discover}. With the Discover stage now occurring before the Winnow and Cast stages, we find that it differs in who the visualization researchers learn about the problem from. Rather than asking or observing front-line analysts to understand pre-existing work tasks, the researchers must work with presentation experts and content-specific literature to understand the \change{concept} to be \change{communicated}. The researchers then abstract the concept into a set of visualization tasks which users can conduct to help them understand the concept.}

\textbf{Winnow}. \change{Since concept discovery occurs before connecting with collaborators, winnowing should focus on finding suitable collaborators for the chosen concept. While this project began as a collaboration between the vis team and the GER team, our collaboration with local museums was not confirmed until a later stage, and we chose to work with these museums due to the applicability of our chosen concept to their educational goals.}

\textbf{Cast}. We found differences between the roles described in the DSM and those involved in our project. The largest difference is with the front-line analyst, a role which did not exist in our project. Instead, we have \textbf{viewers}, describing the museum visitors, who seek to learn about previously-gleaned insights rather than analyze data for new ones and who are the targets of the exposition in the exhibit or exercise. By definition, viewers are not domain experts as they are described in the DSM. Additionally, where the DSM implies that all roles outside of the researchers are held by members of a collaborating institution, the users in our design study were instead the individuals who were served by the institutions. This level of indirection meant that target users were further removed from the other roles.

Museum staff and the GER team were both domain experts, with the museum staff mostly providing expertise on presentation methods in the museum context and the GER team mostly providing expertise on the education methods for the content of the project, deep time. To differentiate between these groups, we label the museum staff as \textbf{presentation experts} and the GER team as \textbf{content experts}. It is possible that one individual could hold both roles, such as a university class instructor who is an expert in both their topic and mode of presentation.

In our project, the museum staff held the gatekeeper role, however without the same level of power as described in the DSM. While they did have the power to approve or block deployment of the exhibit at their institution, they did not have the power to block deployment in other locations or to block access to data.

\change{\textbf{Acquire}. Similar to Oppermann et al.'s data-first design study methodology, we include an explicit Acquire stage, however it occurs after the design stage in our process. The DSM warns against beginning a project where data acquisition is uncertain during the winnow stage. It argues that ``real'' data must exist and be accessible to the visualization researchers. In contrast, we focused on refining our concept to communicate in early stages, finding data to support us in accomplishing this goal in a later stage. In fact, DeLVE's datasets did not exist prior to the project, and museum staff curated them specifically for this exhibit, using specifications for data acquisition that were informed by the design rather than the other way around. While it is important to consider early on whether the data one's design relies on will be available, acquiring or curating it must wait until the Design stage so that the final deployments use data which adequately fits the data abstraction.}

\textbf{Deploy}. Finally, given the difference in users, validation of presentation-focused design studies must differ. The DSM reports that case studies are the most common form of design study validation, but this method is likely insufficient for presentation-focused scenarios. Learner groups in most educational environments are an extremely large and diverse group. While a team of data analysts can simply confirm the usefulness of a visualization tool for a specific analysis task, researchers conducting presentation-focused design studies will likely need to conduct more in-depth field studies with diverse sets of participants to validate the project.

\section{Conclusion and Future Work}
\label{sec:conclusion-and-future-work}

In this paper, we present a task abstraction for supporting proportional reasoning through comparing varied-magnitude time periods. We provide a set of requirements for exhibits in museums, four of them very general and one more tied to the specific learning goal of proportional reasoning. We also identify a data abstraction to characterize how datasets must be curated to support this learning goal. We present the design and implementation of DeLVE, including \change{our Connected Multi-Tier Range idiom, our proposed visualization technique}. We deploy DeLVE in three museums, which entailed achieving approval for deployment and museum-staff curation of datasets from three different institutions. We conduct and report on an observational study and a trace log study to understand user interaction with DeLVE and how it differs between varying museum contexts, including two different locations within the same institution. Finally, we reflect on the project and discuss generalizeable insights for visualization-based museum exhibit design and \change{concept-first} visualization design study methodology.

The GER team has committed to conducting a lab study on DeLVE's ability to improve users' deep time knowledge and proportional reasoning ability. Given the long-term commitments to host DeLVE at M-Bio and M-Sci, future work could study participant behaviour in the exhibit's finalized locations. \change{Since we designed the CMTR idiom for supporting proportional reasoning for deep time, it may be less applicable to other domains.} Other future work on DeLVE \change{and the CMTR idiom} could investigate its application to other domains that could include larger time scales than Earth's formation, like astronomy, which is still related to deep time, or those that use shorter time scales than deep time, like patient health data on the scales of days, weeks, months, and years, which is not deep time but still exponentially increasing. Future studies could also investigate using DeLVE for visualizing scale differences in physical space rather than time, or in formal learning environments such as university classrooms or labs. Finally, future work could evaluate the CMTR idiom for a wider set of usage environments, beyond the museum setting of the DeLVE exhibit. 

\section*{Supplemental Materials}
\label{sec:supplemental_materials}

All supplemental materials are available on OSF at~\url{https://osf.io/z53dq}, released under a CC-BY-4.0 license.
We provide a document with additional details on the project including the prototype evolution, deployments, result figures, and scripts and protocols. We also provide the datasets used for DeLVE during this project, the transcribed data from our observational study, DeLVE's logs, and transcriptions of the expert interviews and workshops. The video showing DeLVE's look and feel is also available at~\url{https://youtu.be/jAIgn3n_-Ss}, DeLVE's source code and instructions for running it are also available at~\url{https://github.com/marasolen/deeptime}. Finally, a live demo of DeLVE is available at~\url{https://deeptime.cs.ubc.ca/}. 

\acknowledgments{
This study was approved by the UBC BREB (H22-02220). Our work was supported in part by UBC STAIR and NSERC DG RGPIN-2014-06309. We thank our museum staff collaborators Jackie Chambers and Derek Tan of the Beaty Biodiversity Museum (M-Bio); Kirsten Hodge of the Pacific Museum of Earth (M-Geo); and Tom Cummins, Michael Fairchild-Simms, and Kristin Lee of Science World (M-Sci). We thank Steve Kasica, Francis Nguyen, Matt Oddo, and Ryan Smith of the UBC InfoVis group, and Ben Shneiderman, for their helpful feedback.
}

\bibliographystyle{abbrv-doi-hyperref}
\bibliography{main}

\end{document}